\documentstyle[psfig,10pt,aaspp4]{article}

\def\Nwidths{522}
\def\Nmem{453}
\def\Ntot{341}
\def\Nopt{324}
\def\NHI{27}
\def\kms{km~s$^{-1}$}
\def\hal{H$\alpha$}
\def\be{\begin{equation}}
\def\ee{\end{equation}}
\def\about{$\sim$}
\def\etal{{\it et al.}}
\def\h{$h^{-1}$}
\def\deg{$^{\circ}$}

\def\rA{$R_{\rm A}$}
\def\Ropt{$R_{\rm opt}$}
\def\r83{$r_{83}$}
\def\Wobs{$W_{\rm obs}$}

\def\HI{\ion{H}{1}}

\def\NII{[\ion{N}{2}]}
\def\SII{[\ion{S}{2}]}

\begin{document}
\hskip 3.5in{\hskip 10pt \date{5 November 1997}}
\title{SEEKING THE LOCAL CONVERGENCE DEPTH. IV. TULLY-FISHER OBSERVATIONS OF 35 
ABELL CLUSTERS}
\altaffiltext{1}{Now at IPAC, California Institute of Technology 100-22, Pasadena, CA 91125}
 
\author {DANIEL A. DALE,\altaffilmark{1} RICCARDO GIOVANELLI, 
MARTHA P. HAYNES, }
\affil{Center for Radiophysics and Space Research and National Astronomy and 
Ionosphere Center, Cornell University, Ithaca, NY 14853}

\author {EDUARDO HARDY}
\affil{National Radio Astronomy Observatory, Casilla 36-D, Santiago, Chile}

\author {LUIS E. CAMPUSANO}
\affil{Observatorio Astron\'{o}mico Cerro Cal\'{a}n, Departamento de 
Astronom\'{\i}a, Universidad de Chile, Casilla 36-D, Santiago, Chile}

\begin{abstract}
We present Tully-Fisher observations for 35 rich Abell clusters of galaxies.  Results from $I$ band photometry and optical rotation curve work comprise the bulk of this paper.  This is the third such data installment of an all--sky survey of 52 clusters in the distance range \about 50 to 200\h\ Mpc.  The complete data set provides the basis for determining an accurate Tully-Fisher template relation and for estimating the amplitude and direction of the local bulk flow on a 100\h\ Mpc scale.
\end{abstract}


\section {INTRODUCTION}

Observational efforts to establish a kinematical reference frame within \about 
100\h\ Mpc (where $H_\circ= 100h$ \kms\ Mpc$^{-1}$) have facilitated detailed studies of large scale velocity fields.  The  Tully-Fisher (TF) programs of Courteau \etal\ (1993), Mathewson, Ford \& Buckhorn (1992) and Giovanelli \etal\ (1998a,b), to name a few, have enabled local bulk flow estimates using the peculiar velocities for several thousand field and cluster spiral galaxies out to about 9,000 \kms.  It is difficult to achieve a similar sampling rate per unit volume, or resolution, on larger distance scales.  On the other hand, kinematic zero-points estimated from peculiar velocity surveys on large scales are less apt to be contaminated by the vagaries of cosmic motions: the relative contribution of peculiar velocities to redshifts diminishes on average with increasing distance.  Kinematical reference frames are thus less certain if derived from ``nearby'' studies.

This drawback of proximate pursuits is accentuated in light of claims made by Scaramella, Vettolani \& Zamorani (1994), Tini-Brunozzi \etal\ (1995), and Branchini, Plionis \& Sciama (1996).  They suggest the local convergence depth, i.e. the distance out to which significant contributions to the Local Group motion originate, may be as large as 180\h\ Mpc.  It is therefore important to probe the velocity field over a similar distance scale.  Lauer \& Postman (1994; LP) accomplished this by using the Hoessel relation to determine  redshift-independent distances to the brightest galaxies in the nearest 119 Abell clusters.  They found the surprising result that the volume of space within \about 100\h\ Mpc could be moving at \about 700$\pm$200 \kms\ with respect to the inertial reference frame in which the cosmic microwave background (CMB) radiation dipole is null.  Their claim has been contested over the past few years by a variety of results on both theoretical (e.g. Feldman \& Watkins 1994; Gramann \etal\ 1995; Strauss \etal\ 1995; Moscardini \etal\ 1996; M\"{u}ller \etal\ 1998) and observational grounds (Riess, Press \& Kirshner 1995; Giovanelli \etal\ 1998a,b).  Due to limitations in the depth and sampling of the available data, a conclusive experimental test has however been difficult to execute.  Results from more recent work at larger distances are beginning to appear in the literature, e.g. Giovanelli 1998c, Dale \etal\ 1999; Hudson \etal\ 1999; Willick 1999; here we present data relevant to one of these efforts. 

We expand upon the data of Giovanelli \etal\ 1997a,b (G97a,b) by obtaining TF measurements for an all--sky survey of 52 clusters in the redshift range 5000 $\lesssim cz \lesssim$ 20,000 \kms\ (hereafter the ``SCII'' sample).  The benefits of such an enterprise are twofold.  First, the two data sets yield a highly accurate TF template.  The proximity of the extant G97a,b sample allows a wide range of galactic properties to be observed and is thus well-suited for exploring environmental dependencies and for determining the slope of the TF relation.  Conversely, our relatively more distant sample allows an accurate kinematic zero-point estimation, as explained above.  Second, the combined data set covers a much larger volume than that of G97a,b alone.  This enables dipole motion measurements over scales large enough to effectively test the convergence depth of the Local Universe.

In order to provide public access to the data on the shortest possible time 
scale, we have presented results of our survey in installments, as we progressed in the data reductions for sizable fractions of the cluster set.  Dale \etal\ (1997 \& 1998; hereafter Papers I and II) presented TF measurements for 84 and 90 galaxies in the fields of seven and ten Abell clusters, respectively.  In this paper, we present the project's last data installment: TF measurements for \Ntot\ galaxies in the fields of 35 Abell clusters.  The data for the entire series of papers can be obtained by contacting the first author.  The following section reviews the imaging and spectroscopic observations for this data installment.  Section 3 presents the relevant TF data and highlights the general properties of the entire data set. 

\section {OBSERVATIONS}					

Imaging for this project began in October 1994 and the spectroscopy measurements began in December 1995.  We have obtained 489 TF measurements via optical imaging and optical spectroscopy.  Including 21 cm data taken earlier by R.G. and M.P.H. for approximately three dozen other galaxies, we have obtained \Nwidths\ velocity widths in the fields of 52 Abell clusters, with \Nmem\ of these belonging to actual cluster members.  Paper V of this series, a companion paper in this edition of the {\it Journal}, presents the TF template relation derived from this set of 52 clusters, along with their peculiar velocities.

\subsection {$I$ Band Imaging}

\begin{figure}[!ht]
\centerline{\psfig{figure=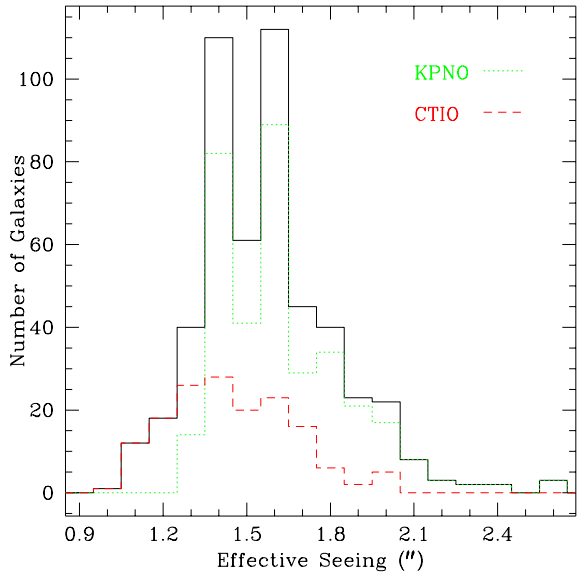,width=3.0in,bbllx=15pt,bblly=140pt,bburx=548pt,bbury=673pt}}
\caption[Effective Seeing Distribution]
{\ The distribution of effective seeing for the galaxy observations.  The dotted line is for KPNO observations, the dashed line is for CTIO observations, and the solid line is the sum of the two distributions.}
\label{fig:seeing}
\end{figure}

All photometric observations were done in the $I$ band, matching the filter used for the G97a,b data.  The imaging for the clusters in this paper was carried out at the KPNO and CTIO 0.9 m telescopes.  We do not review here the procedure used to take and reduce the photometry data, but we instead refer the interested reader to the relevant sections in Papers I and II.  Table \ref{tab:image_runs} highlights pertinent information for all of the runs associated with this work.  
\begin{table} 
\dummytable\label{tab:image_runs}
\end{table}
The seeing conditions for the entire data sample (i.e. including Paper I and II data) is displayed in Figure \ref{fig:seeing}.  
The average FWHM seeing for the images was 1.6\arcsec\ $\pm$ 0.3\arcsec\ at KPNO (with 0.3\arcsec\ being the 1$\sigma$ dispersion in the seeing) and 1.4\arcsec\ $\pm$ 0.2\arcsec\ at CTIO; however, the nights with the best seeing conditions were preferentially devoted to the more distant clusters.  The majority of the data presented here were taken in good photometric conditions, for which the photometric zero-point calibration could be determined with an accuracy of 0.02 mag or better.  In a minority of cases (2\%), photometric conditions were of inferior quality.  However, since each cluster field required several adjacent images, the overlap of contiguous images by approximately 10\% of the sky area of each allowed photometric calibration of the frames taken in poor photometric conditions.  Provided that fluxes of at least 12 field stars could be measured in the overlap region of 
high photometric quality, calibration to the \about\ 0.03 mag level could be guaranteed.

\subsection {Optical Spectroscopy}
\label{sec:spec}

Rotational velocity widths for this sample of cluster galaxies were extracted from long--slit spectra obtained at the Palomar Observatory 5 m telescope and the CTIO 4 m telescope.  Table 2 lists the observing runs and summarizes their technical setups.  
\begin{table}
\dummytable\label{tab:spec_runs}
\end{table}
Again, the interested reader can find relevant details on the spectroscopy data in Papers I and II.  A review of the entire data set reveals that we detected line emission in 582 out of 897 (65\%) targeted galaxies observed in these 52 clusters.

Rotation curves, redshifts, and velocity widths are extracted from the spectroscopy employing the same techniques used for Papers I and II.  We refer redshifts (c$z$) to the kinematical center of the rotation curve, merely the average of the 10\% and 90\% velocities (an N\% velocity is greater than N\% of all the velocity data points).  We also present raw velocity widths as the difference between these two velocities, i.e. c$z \equiv (V_{\rm 90\%} + V_{\rm 10\%}$)/2 and \Wobs\ $\equiv (V_{\rm 90\%} - V_{\rm 10\%})$.  For galaxies with a rotation curve that extends to \Ropt, the distance along the disk major axis to the isophote containing 83\% of the $I$ band light (cf. Persic, Salucci \& Stel 1996; G97a), \Wobs\ matches well the velocity width at \Ropt.  Giovanelli \etal\ (1999; G99) compare this method for estimating rotational velocity widths with other measures, including those that use the velocity width measured near two exponential disk scale lengths (e.g. Courteau \& Rix 1999; Willick 1999).  They find that using the width at a fixed radial distance from the galaxy center far enough in the periphery of the disk, such as \Ropt, minimizes distortions produced by dynamics and internal extinction.

Observed rotational velocity widths suffer from a number of effects that combine to misrepresent \Wobs\ as a fair estimate of the full rotational speed within a disk.  Papers I and II describe our approaches to correcting for the following effects: disk inclination $i$; the underestimate of $i$ in poor seeing conditions; the cosmological stretching of wavelengths; and the radial undersampling of the spiral disk.  (Approximately 40\% of our rotation curves do not extend to \Ropt.  For these galaxies we extrapolate the observed rotation curves to \Ropt.)  We have referred to the correction for the latter effect as $\Delta_{\rm sh}$ since the correction is a function of the shape of the rotation curve.  The correction rarely exceeds 0.1 \Wobs.  The velocity width that is corrected for these effects is referred to as
\be
W_{\rm cor} = {{W_{\rm obs} + {\Delta}_{\rm sh}} \over {(1+z)\sin i}}.
\ee
Other factors do, however, bias observed velocity widths, including disk opacity, spectral profile smearing within the spectrograph slit, and the incorrect positioning of the slit with respect to the disk's major axis.  A full discussion of these effects is given in G99.  For consistency with Papers I and II, the data presented here do not include corrections for disk opacity and slit-related effects.  The latter are, however, relatively small, typically of order 2\%.  They are fully taken into account when the data are used for astrophysical inferences.

In addition to the \Nopt\ rotation curves presented here, we also use 21 cm measurements for \NHI\ galaxies, taken with the Arecibo telescope with a spectral resolution of 8 \kms\ as described in Haynes \etal\ (1997).  A typical signal to noise ratio for these observations was 10; errors in the observed velocity widths are of order 15 \kms.  G99 make a comparison between \HI\ and optical widths and they show that our method for estimating optical velocity widths closely recovers widths obtained from 21 cm single-dish data; they explain our simple method for cross-calibrating the two samples of velocity widths.

\section {DATA}					

Table 3 lists the main parameters of the clusters.  
\begin{table}
\dummytable\label{tab:clusters}
\end{table}
Standard names are listed in column 1.  The fields of two of these systems, A2877 and A1736, each contain two separate clusters.  Adopted coordinates of the cluster center are listed in columns 2 and 3, for the epoch 1950; they are obtained from Abell, Corwin, and Olowin (1989).  For all the clusters we derived a new systemic velocity, combining the redshift measurements available in the NED\footnote{The NASA/IPAC Extragalactic Database is operated by the Jet Propulsion Laboratory, California Institute of Technology, under contract with the National Aeronautics and Space Administration.} database with our own measurements.  These newly determined velocities are listed in columns 4 and 5, in the heliocentric and in the CMB reference frame (Kogut \etal\ 1993), respectively.  We list the number of cluster member redshifts used in determining systemic velocities in column 6.  An estimated error for the systemic velocity is parenthesized after the heliocentric figure.  Spherical and Cartesian supergalactic coordinates are given in columns 7 and 8, and in columns 9--11, respectively.

\begin{figure}[!ht]
\centerline{\psfig{figure=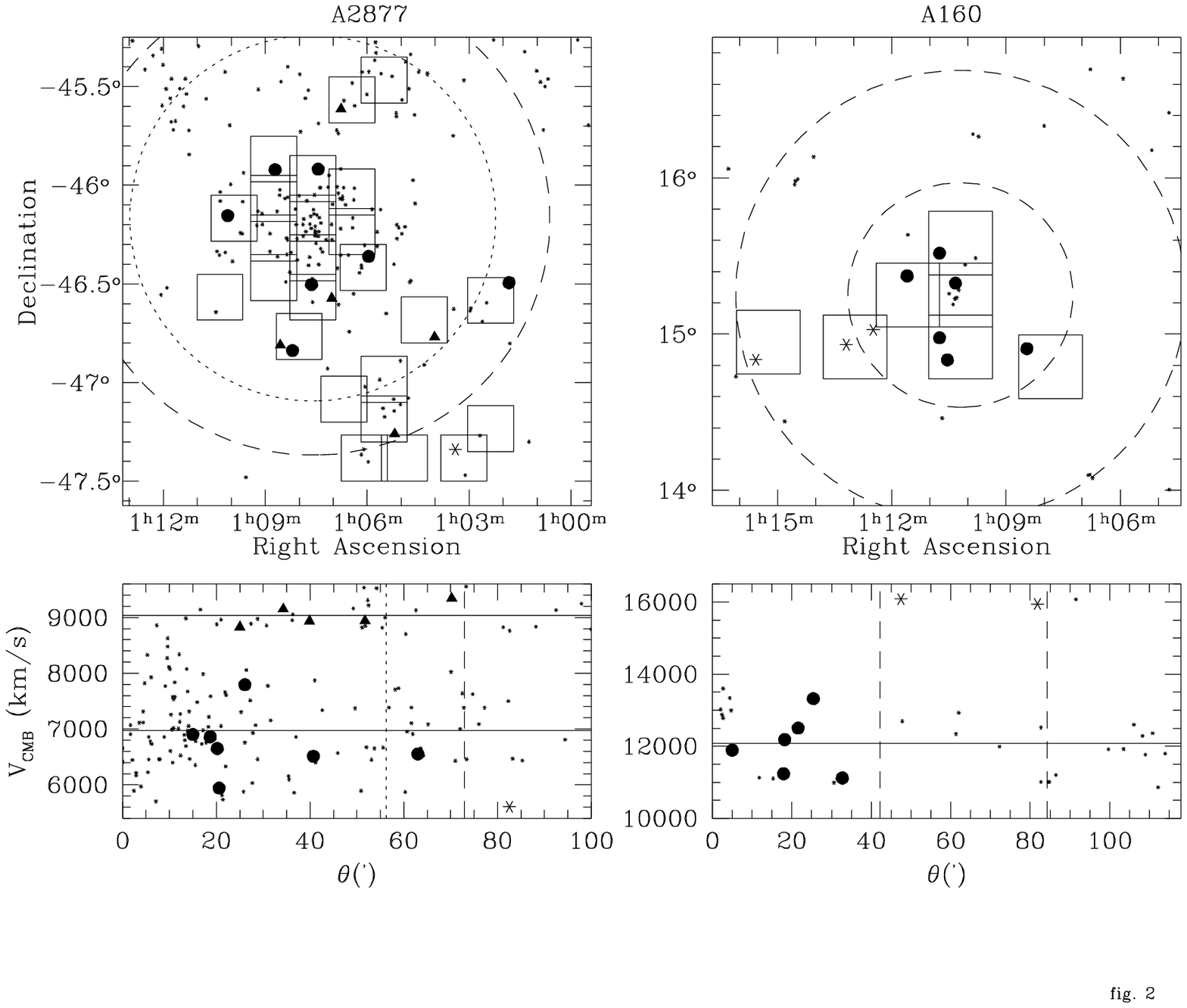,width=5.0in,bbllx=35pt,bblly=213pt,bburx=578pt,bbury=617pt}}
\caption[]{Sky and velocity distribution of galaxies in the clusters Abell 2877, Abell 2877b, and Abell 160.  Circles represent cluster members with measured photometry and widths; if unfilled, widths are poorly determined.  For A2877b,
filled triangles identify cluster members.  Asterisks identify foreground and background galaxies and dots give the location of galaxies with known redshift, but lacking accurate width and/or photometry.  Large square boxes indicate outlines of imaged fields with the 0.9 m telescope.  Vertical dashed lines in the lower panels indicate 1\rA\ (and 2\rA\ for A160); a dotted line is used for A2877b.  The upper panels contain circles of radius 1\rA\ (and 2\rA\ for A160).
\label{a2877}}
\end{figure}

Figures \ref{a2877}--\ref{a4038} show the distribution of the galaxies in each cluster.  The top panel in each of these figures displays the spatial location of:  the outline of the fields imaged (large squares), cluster members (circles --- those with poor/unusable velocity widths are left unfilled), background or foreground objects (asterisks), and galaxies with known redshift but without reliable widths (dots).  Circles of 1 and 2 Abell radii, \rA, are drawn as dashed lines, if the area displayed is large enough.  If no dashed circle is drawn, \rA\ is larger than the figure limits.  We also plot the radial CMB velocity as a function of angular distance from the cluster center in the lower panel of each figure.  A dashed vertical line is drawn at 1 \rA.  The combination of the sky and velocity plots is used to gauge cluster membership for each galaxy.  
\figcaption[]{Sky and velocity distribution of galaxies in the clusters Abell 260 and Abell 3266.  Filled circles, unfilled circles, asterisks, dots, large squares and dashed lines and circles follow the same convention as in Figure \ref{a2877}.}

\figcaption[]{Sky and velocity distribution of galaxies in the clusters Abell 496 and Abell 3407.  Filled circles, unfilled circles, asterisks, dots, large squares and dashed lines and circles follow the same convention as in Figure \ref{a2877}.}

\figcaption[]{Sky and velocity distribution of galaxies in the clusters Abell 634 and Abell 671.  Filled circles, unfilled circles, asterisks, dots, large squares and dashed lines and circles follow the same convention as in Figure \ref{a2877}.}

\figcaption[]{Sky and velocity distribution of galaxies in the clusters Abell 754 and Abell 779.  Filled circles, unfilled circles, asterisks, dots, large squares and dashed lines and circles follow the same convention as in Figure \ref{a2877}.}

\figcaption[]{Sky and velocity distribution of galaxies in the clusters Abell 957 and Abell 1177.  Filled circles, unfilled circles, asterisks, dots, large squares and dashed lines and circles follow the same convention as in Figure \ref{a2877}.}

\figcaption[]{Sky and velocity distribution of galaxies in the clusters Abell 1213 and Abell 1314.  Filled circles, unfilled circles, asterisks, dots, large squares and dashed lines and circles follow the same convention as in Figure \ref{a2877}.}

\figcaption[]{Sky and velocity distribution of galaxies in the clusters Abell 3528, Abell 1736, and Abell 1736b.  For A3528 and A1736, filled circles, unfilled
circles, asterisks, dots, large squares and dashed lines and circles follow the
same convention as in Figure \ref{a2877}.  For A1736b, filled triangles identify cluster members and 1 \rA\ is indicated by the dotted line in the lower panel and by the dotted circle of radius 1 \rA\ in the upper panel.}

\figcaption[]{Sky and velocity distribution of galaxies in the clusters Abell 3558 and Abell 3566.  Filled circles, unfilled circles, asterisks, dots, large squares and dashed lines and circles follow the same convention as in Figure \ref{a2877}.}

\figcaption[]{Sky and velocity distribution of galaxies in the clusters Abell 3581 and Abell 2022.  Filled circles, unfilled circles, asterisks, dots, large squares and dashed lines and circles follow the same convention as in Figure \ref{a2877}.}
 
\figcaption[]{Sky and velocity distribution of galaxies in the clusters Abell 2040 and Abell 2063.  Filled circles, unfilled circles, asterisks, dots, large squares and dashed lines and circles follow the same convention as in Figure \ref{a2877}.}
 
\figcaption[]{Sky and velocity distribution of galaxies in the clusters Abell 2147 and Abell 2151.  Filled circles, unfilled circles, asterisks, dots, large squares and dashed lines and circles follow the same convention as in Figure \ref{a2877}.}
 
\figcaption[]{Sky and velocity distribution of galaxies in the clusters Abell 2256 and Abell 3656.  Filled circles, unfilled circles, asterisks, dots, large squares and dashed lines and circles follow the same convention as in Figure \ref{a2877}.}
 
\figcaption[]{Sky and velocity distribution of galaxies in the clusters Abell 3667 and Abell 3716.  Filled circles, unfilled circles, asterisks, dots, large squares and dashed lines and circles follow the same convention as in Figure \ref{a2877}.}
 
\figcaption[]{Sky and velocity distribution of galaxies in the clusters Abell 2572 and Abell 2589.  Filled circles, unfilled circles, asterisks, dots, large squares and dashed lines and circles follow the same convention as in Figure \ref{a2877}.}
 
\figcaption[]{Sky and velocity distribution of galaxies in the clusters Abell 2593 and Abell 2657.  Filled circles, unfilled circles, asterisks, dots, large squares and dashed lines and circles follow the same convention as in Figure \ref{a2877}.}
 
\figcaption[]{Sky and velocity distribution of galaxies in the cluster Abell 4038.  Filled circles, unfilled circles, asterisks, dots, large squares and dashed
lines and circles follow the same convention as in Figure \ref{a2877}.
\label{a4038}}

We separate photometric data and spectroscopic data into two tables.  Table 4 lists the spectroscopic properties and Table 5 gives the pertinent photometric results.  

Entries in the tables are sorted first by the Right Ascension of each cluster, 
and within each cluster sample by increasing galaxy Right Ascension.  The listed parameters for Table 4 are:

\noindent 
Col. 1: identification names corresponding to a coding number in our private
database, referred to as the Arecibo General Catalog, which we maintain for easy reference in case readers may request additional information on the object.

\noindent
Cols. 2 and 3: Right Ascension and Declination in the 1950.0 epoch.  Coordinates
have been obtained from the Digitized Sky Survey and are accurate to
$<$ 2\arcsec.

\noindent
Cols. 4 and 5: the galaxy radial velocity as measured in the heliocentric and
CMB reference frame (Kogut \etal\ 1993).  Errors are parenthesized: e.g. 
13241(08) implies 13241$\pm$08.

\noindent
Col. 6: the raw velocity width in \kms. Measurement of optical widths are 
described in Section \ref{sec:spec} and Section 2.2 of Paper II; 21 cm line widths are denoted with a dagger and refer to values measured at a level of 50\% of the profile horns.

\noindent
Col. 7: the velocity width in \kms\ after correcting for rotation curve shape, the cosmological stretch of the data and, for 21 cm data, signal to noise effects, interstellar medium turbulence, and instrumental and data processing broadening; details on the adopted corrections for optical and 21 cm data are given in Section 2 of Paper II and G97a, respectively.

\noindent
Col. 8:  the corrected velocity width in \kms\ converted to an edge--on perspective.

\noindent
Col. 9: the adopted inclination $i$ of the plane of the disk to the line of 
sight, in degrees, (90$^\circ$ corresponding to edge--on perspective); the 
derivation of $i$ and its associated uncertainty are discussed in Section 4 of 
Paper I.

\noindent
Col. 10: the logarithm in base 10 of the corrected velocity width (value in 
column 7), together with its estimated uncertainty between brackets.  The 
uncertainty takes into account both measurement errors and uncertainties arising from the corrections.  The format 2.576(22), for example, is equivalent to 
2.576$\pm$0.022.

The position angle adopted for the slit of each spectroscopic observation is 
that given in column 4 of Table 5.  The first column in Table 5 matches that of 
Table 4.  The remaining listed parameters for Table 5 are:

\noindent
Col. 2: morphological type code in the RC3 scheme (de Vaucouleurs \etal\ 1991), where code 1 corresponds to Sa, code 3 to Sb, code 5 to Sc, code 7 to Sd, etc.  When the type code is followed by a ``B'', the galaxy disk has an identifiable bar.  We assign these codes after visually inspecting the CCD $I$ band images and after noting the value of $R_{75}/R_{25}$, where $R_X$ is the radius containing X\% of the $I$ band flux.  This ratio is a measure of the central concentration of the flux which was computed for a variety of bulge--to--disk ratios.  Given the spatial resolution of the images, some of the inferred types are rather uncertain; uncertain types are followed by a colon.

\noindent
Col. 3: the angular distance $\theta$ in arcminutes from the center of each cluster.

\noindent
Col. 4: position angle of the disk major axis, also used for spectrograph slit positioning (North to East: 0$^{\circ}$ to 90$^{\circ}$).

\noindent
Col. 5: observed ellipticity of the disk.
 
\noindent
Col. 6: ellipticity corrected for seeing effects as described in Section 5 of 
Paper I, along with its corresponding uncertainty.  The format 0.780(16), for 
example, is equivalent to 0.780$\pm$0.016.

\noindent
Col. 7: $I$ band surface brightness of the disk at zero radius, as extrapolated from the fit to the one-dimensional disk surface brightness profile.

\noindent
Col. 8: the $I$ band (exponential) disk scale length.

\noindent
Col. 9: the distance along the major axis to the isophote containing 83\% of the $I$ band flux.

\noindent
Col. 10: isophotal radius along the major axis where the $I$ band surface brightness equals 23.5 mag sec$^{-2}$.

\noindent
Col. 11: $I$ band apparent magnitude within the 23.5 mag sec$^{-2}$ isophote.

\noindent
Col. 12: the measured $I$ band magnitude, extrapolated to eight face-on disk scale lengths, assuming that the surface brightness profile of the disk is well described by an exponential function.

\noindent
Col. 13: the $I$ band apparent magnitude, to which $k$-term, Galactic and internal extinction corrections were applied.  Galactic extinction estimates (for all SCII clusters) have been updated according to Schlegel, Finkbeiner \& Davis (1998); details on the adopted internal extinction corrections are given in Section 2.1 of Paper II. 

\noindent
Col. 14: the $I$ band absolute magnitude, computed assuming that the galaxy is at the  distance indicated either by the cluster redshift, if the galaxy is a cluster member within 1 \rA\ of the cluster center, or by the galaxy redshift if not.  The calculation assumes $H_\circ = 100h$ \kms\ Mpc$^{-1}$, so the value listed is strictly $M_{\rm cor} - 5\log h$.  In calculating this parameter, radial velocities are expressed in the CMB frame and uncorrected for any cluster peculiar motion.  The uncertainty on the magnitude, indicated between brackets in hundredths of a mag, is the sum in quadrature of the measurement errors and the estimate of the uncertainty in the corrections applied to the measured parameter.

When an asterisk appears at the end of the record, a  detailed comment is given 
for that particular object.  Because of the length and number of these comments, they are not appended to the table but included in the text as follows.  Note
that a record is flagged in both Tables 4 and 5, independently on whether the
comments refer only to the photometry, only to the spectroscopy, or both.  Acronyms and abbreviations used in the notes include: RC for rotation curve; PA for position angle; Pal I for observations made at the Palomar 5 m telescope by a previous effort (Vogt, Haynes, Herter \& Giovanelli).

\noindent {\bf A2877:}\\ \small
\noindent 20791: Foreground galaxy.\\
\noindent 410675: Note low $i$.\\
\noindent  20893: Asymmetric $I$ band profile.

\normalsize
\noindent {\bf A2877b:}\\ \small
\noindent  20797: Note low $i$.\\
\noindent  20845: Note low $i$; ellipticity uncertain.

\normalsize
\noindent {\bf A160:}\\ \small
\noindent 110104: Only 21 cm width available.\\
\noindent 110968: Unusual spiral arm pattern.\\
\noindent 110975: Background galaxy; 5 minute integration.\\
\noindent 110976: Background galaxy.\\
\noindent 110977: Background galaxy.

\normalsize
\noindent {\bf A260:}\\ \small
\noindent 110558: Marginal quality 21 cm width also available.\\
\noindent 110805: Mostly bulge RC; \NII\ patch on receding side of RC for $r \gtrsim$2\arcsec; companion 45\arcsec\ to East; unfit for TF use.\\
\noindent 110937: Background galaxy; 5 minute integration.

\normalsize
\noindent {\bf A3266:}\\ \small
\noindent 440559: Note low $i$; uncertain ellipticity; unfit for TF use; center-of-light peak used for RC spatial center.

\normalsize
\noindent {\bf A496:}\\ \small
\noindent 440516: Atypical emission characteristics on receding side of RC for both
\hal\ and \SII\ lines.\\
\noindent 440519: Rising RC.

\normalsize
\noindent {\bf A3407:}\\ \small
\noindent 470039: Uncertain ellipticity; image partly truncated by frame edge; complex velocity profile; unfit for TF use.\\
\noindent 470051: Background galaxy.\\
\noindent 470058: Disk warp?

\normalsize
\noindent {\bf A634:}\\ \small
\noindent 180807: Foreground galaxy; slit PA off 9\deg\ from disk major axis PA.\\ 
\noindent 180715: Mostly bulge RC; unfit for TF use.

\normalsize
\noindent {\bf A671:}\\ \small
\noindent 180287: Uncertain RC extrapolation.\\
\noindent 180802: Background galaxy; 5 minute integration.\\
\noindent 180803: Background galaxy.

\normalsize
\noindent {\bf A754:}\\ \small
\noindent 490235: Foreground galaxy; 5 minute integration.\\
\noindent 490249: 5 minute integration.

\normalsize
\noindent {\bf A779:}\\ \small
\noindent   4865: Pal I RC; marginal quality 21 cm width also available.\\
\noindent   4926: Pal I RC; 21 cm width also available.\\
\noindent 190187: Only 21 cm width available.\\
\noindent   4941: Pal I RC; mildly discrepant (5\%) optical and 21 cm widths; companions.\\
\noindent   4949: Pal I RC; discrepant (10\%) optical and 21 cm widths.\\
\noindent 190224: Pal I RC; 21 cm width preferred.\\
\noindent 190225: Only 21 cm width available; star nearby.\\
\noindent   4960: Only 21 cm width available; Note low $i$.\\
\noindent 190243: 21 cm width also available.\\
\noindent 190203: Pal I RC.\\
\noindent 190252: Only 21 cm width available.\\
\noindent   5014: Only 21 cm width available.\\
\noindent 190765: Only 21 cm width available.\\
\noindent 190293: Only 21 cm width available.\\
\noindent   5041: Only 21 cm width available.

\normalsize
\noindent {\bf A957:}\\ \small
\noindent 500048: 1\arcsec\ slit used.\\
\noindent 500344: Background galaxy.\\
\noindent 500345: Background galaxy; 5 minute integration.

\normalsize
\noindent {\bf A1177:}\\ \small
\noindent 211452: Background galaxy; incorrect PA used.\\
\noindent 211453: Foreground galaxy; asymmetric $I$ band profile; possible stripped disk.\\
\noindent 211459: Foreground galaxy.\\
\noindent 211499: Rising RC.\\
\noindent 211500: \NII\ RC (\hal\ RC does not extend as far radially).

\normalsize
\noindent {\bf A1213:}\\ \small
\noindent 211561: Background galaxy; 5 minute integration.\\
\noindent 211564: \NII\ RC (\hal\ RC unavailable).\\
\noindent   6292: Interacting? Distorted spiral structure; unfit for TF use.

\normalsize
\noindent {\bf A1314:}\\ \small
\noindent 211621: Rising RC.\\
\noindent 211634: Asymmetric $I$ band profile.\\
\noindent 211641: Dust lane.

\normalsize
\noindent {\bf A3528:}\\ \small
\noindent 520509: Background galaxy.\\
\noindent 520227: Background galaxy.\\
\noindent 520482: Background galaxy; 5 minute integration; rising RC.\\
\noindent 520535: Background galaxy.\\
\noindent  28988: Foreground galaxy; 5 minute integration.\\
\noindent 520530: Background galaxy.

\normalsize
\noindent {\bf A1736:}\\ \small
\noindent  29640: Star nearby; low S/N RC data.

\normalsize
\noindent {\bf A1736b:}\\ \small
\noindent 530430: Uncertain RC extrapolation; unfit for TF use.\\
\noindent 530432: Not kinematically centered; \Wobs\ is twice the rotational velocity of the receding arm due to lack of emission from approaching arm.

\normalsize
\noindent {\bf A3558:}\\ \small
\noindent 530179: Interacting triplet?\\
\noindent 530448: RC not kinematically centered.\\
\noindent 530456: \NII\ patch for radii $\lesssim$2\arcsec.\\
\noindent 530471: Uncertain ellipticity; warped disk?\\
\noindent 530474: Slit PA off 17\deg\ from disk major axis PA.\\
\noindent 530478: RC not kinematically centered.

\normalsize
\noindent {\bf A3566:}\\ \small
\noindent 530519: Slit PA off 13\deg\ from disk major axis PA.\\
\noindent 530530: Center-of-light peak used for RC center.\\
\noindent 530537: \NII\ patch for radii $\lesssim$1\arcsec.\\
\noindent 530541: \NII\ RC (\hal\ RC does not extend as far radially).

\normalsize
\noindent {\bf A3581:}\\ \small
\noindent 540202: Background galaxy.

\normalsize
\noindent {\bf A2022:}\\ \small
\noindent 250021: Interacting w/250022?\\
\noindent 250022: Interacting w/250021?\\
\noindent 251792: \NII\ RC (\hal\ RC does not extend as far radially); rising RC; center-of-light peak used for RC center.\\
\noindent 251793: Rising RC; slit PA off 9\deg\ from disk major axis PA.\\
\noindent 251802: Background galaxy.

\normalsize 
\noindent {\bf A2040:}\\ \small
\noindent 251754: RC not kinematically centered; \Wobs\ is twice the rotational velocity of the approaching arm due to lack of emission from receding arm.\\
\noindent 250266: \NII\ RC (\hal\ RC does not extend as far radially).\\
\noindent 250374: Uncertain RC extrapolation.

\normalsize 
\noindent {\bf A2063:}\\ \small
\noindent 250443: Pal I RC; 21 cm width used.\\
\noindent 250472: Pal I RC; 21 cm width also available.\\
\noindent 250473: Background galaxy; only 21 cm width available.\\
\noindent 251814: \NII\ RC (\hal\ RC does not extend as far radially); RC still rising.\\
\noindent 251816: Faint dust lane.\\
\noindent 250632: Fair quality 21 cm width also available.\\
\noindent 250641: Bulge off-center; fair quality 21 cm width also available.\\
\noindent 250644: Irregular? Bulge off-center.\\
\noindent 250716: Background galaxy; 5 minute integration.\\
\noindent 250740: Pal I RC; 21 cm width also available.\\
\noindent   9838: Pal I RC.\\
\noindent 250786: \NII\ RC (\hal\ RC does not extend as far radially); fair quality 21 cm width also available.\\
\noindent   9844: Pal I RC; 21 cm width also available.\\
\noindent 250803: Pal I RC; 21 cm width also available.\\
\noindent 250896: Pal I RC; 21 cm width also available.

\normalsize
\noindent {\bf A2147:}\\ \small
\noindent 251498: Only 21 cm width available.\\
\noindent 251753: Note low $i$; slit PA off 15\deg\ from disk major axis PA.\\
\noindent 251400: Pal I RC; confused 21 cm profile also available.\\
\noindent 251424: Pal I RC.\\
\noindent 251503: Pal I RC.\\
\noindent 251436: Pal I RC; incorrect slit PA used.\\
\noindent 251509: Pal I RC.\\
\noindent  10131: Only 21 cm width available.\\
\noindent 251510: Pal I RC; marginal quality 21 cm width also available; average of optical and 21 cm widths used.\\
\noindent 260959: Only 21 cm width available.\\
\noindent 260246: Poor Pal I RC; 21 cm width preferred.  Asymmetric $I$ band profile.\\
\noindent 260963: Only 21 cm width available.\\
\noindent  10287: 21 cm width preferred; Note low $i$; incorrect slit PA used.

\normalsize
\noindent {\bf A2151:}\\ \small
\noindent  10121: Pal I RC; discrepant optical and 21 cm widths; 21 cm width used.\\
\noindent 251442: Only 21 cm width available.\\
\noindent 260091: Pal I RC; uncertain \Wobs.\\
\noindent 260112: Pal I RC; note low $i$; slit PA off 11\deg\ from disk major axis PA.\\
\noindent 260116: Pal I RC.\\
\noindent 260953: Pal I RC; 21 cm width also available.\\
\noindent  10177: Pal I RC; warped disk?  21 cm width also available.\\
\noindent  10180: Pal I RC; slit PA off 11\deg\ from disk major axis PA.\\
\noindent 260146: Pal I RC.\\
\noindent  10190: Pal I RC; 21 cm width preferred.\\
\noindent 260184: Pal I RC; note low $i$; slit PA off 18\deg\ from disk major axis PA.\\
\noindent  10195: Pal I RC.\\
\noindent 260208: Pal I RC; 21 cm width also available; ellipticity uncertain.\\
\noindent 260226: Pal I RC; dust lane; $i$ set to 88\deg.\\
\noindent 260245: Pal I RC.\\
\noindent 260247: Only 21 cm width available.\\
\noindent 260284: Pal I RC; center-of-light peak used for RC spatial center; 21 cm width also available.\\
\noindent 260302: Pal I RC; 21 cm width also available.

\normalsize
\noindent {\bf A2256:}\\ \small
\noindent 261191: Foreground galaxy; 5 minute integration.\\
\noindent 270144: Uncertain ellipticity.\\
\noindent 270148: \NII\ RC (\hal\ RC unavailable); rising RC.\\
\noindent 270385: \NII\ patch for radii $\lesssim$2\arcsec; slit PA off 11\deg\ from disk major axis PA.\\
\noindent 270172: Rising RC.

\normalsize
\noindent {\bf A3656:}\\ \small
\noindent 590070: Background galaxy.\\
\noindent 590071: Background galaxy.\\
\noindent  32903: Star nearby; PA uncertain.\\
\noindent 590083: Background galaxy; star nearby.\\
\noindent 590073: Background galaxy; note low $i$.\\
\noindent 590074: Background galaxy.\\
\noindent  32965: Background galaxy.\\
\noindent 590088: Background galaxy.\\
\noindent 600332: Background galaxy.\\
\noindent 600333: Background galaxy; \NII\ patch for radii $\lesssim$6\arcsec.\\
\noindent 600334: Background galaxy.

\normalsize
\noindent {\bf A3716:}\\ \small
\noindent 600243: Rising RC.\\
\noindent 600251: Foreground galaxy.\\
\noindent  33829: Low S/N on receding side of RC.\\
\noindent 600264: Asymmetric $I$ band profile.\\
\noindent  33844: Rising RC; prominent dust lane.\\
\noindent 600270: RC not kinematically centered.\\
\noindent  33851: \NII\ patch for radii $\lesssim$4\arcsec.\\
\noindent 600289: Star nearby.\\
\noindent 600309: Background galaxy.\\
\noindent  33887: Background galaxy; \NII\ patch for radii $\lesssim$3\arcsec; note low $i$; PA uncertain.

\normalsize
\noindent {\bf A2572:}\\ \small
\noindent 331603: Background galaxy; 5 minute integration.\\
\noindent 331606: Background galaxy; 5 minute integration.

\normalsize
\noindent {\bf A2589:}\\ \small
\noindent 330292: Background galaxy.\\
\noindent 331577: Background galaxy.\\
\noindent 330318: RC not kinematically centered.\\
\noindent 331686: Asymmetric $I$ band profile; uncertain ellipticity.\\
\noindent 331163: Fair quality 21 cm width also available.\\
\noindent 330362: Note low $i$; uncertain RC extrapolation; unfit for TF use.

\normalsize
\noindent {\bf A2593:}\\ \small
\noindent 331695: Irregular?\\
\noindent 331702: Rising RC.\\
\noindent 331726: Note low $i$; \Ropt\ unreliable.\\
\noindent 331727: Asymmetric $I$ band profile.\\
\noindent 331737: \NII\ patch for radii $\lesssim$7\arcsec.

\normalsize
\noindent {\bf A2657:}\\ \small
\noindent 331608: Foreground galaxy; 5 minute integration; member of triplet?\\
\noindent 331611: Companion 40\arcsec\ to East.\\
\noindent 331612: Rising RC.\\
\noindent 330872: \NII\ patch for radii $\lesssim$5\arcsec.

\normalsize
\noindent {\bf A4038:}\\ \small
\noindent  36309: \NII\ patch for radii $\lesssim$5\arcsec.\\
\noindent 630396: Background galaxy; 5 minute integration.\\
\noindent 630397: Background galaxy; Note low $i$.\\
\noindent 630400: \NII\ patch for radii $\lesssim$2\arcsec.\\
\noindent 630408: Note low $i$.\\
\noindent 630414: \NII\ patch for radii $\lesssim$4\arcsec.\\
\noindent  36409: \NII\ patch for radii $\lesssim$1\arcsec.\\
\noindent 630424: Background galaxy; c$z \sim$ 43,000 \kms; rising RC.
\normalsize

\begin{figure}[!ht]
\centerline{\psfig{figure=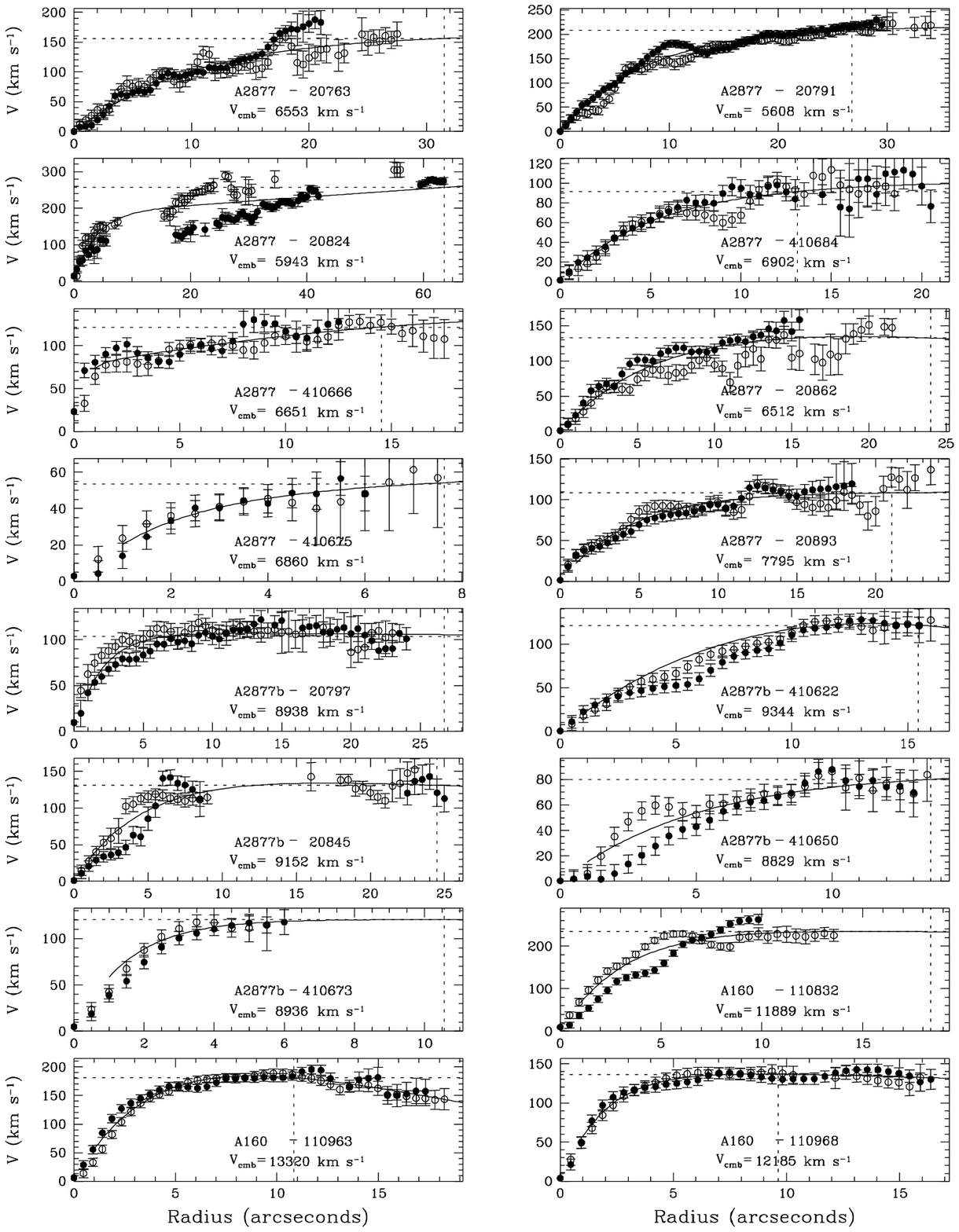,width=5in,bbllx=70pt,bblly=140pt,bburx=514pt,bbury=714pt}}
\caption[]{\hal\ rotation curves for \Nopt\ galaxies (except for galaxies 211500, 211564, 251792, 250266, 251814, 250786, 270148, and 530541, for which the rotation curves are obtained from a \NII\ emission line), folded about the kinematic centers.  Only a portion of the full sample of rotation curves are displayed here; the remainder can be obtained by contacting the first author.  Names of the galaxy and the corresponding parent cluster are given along with the CMB radial velocity.  Two dashed lines are drawn: the horizontal line indicates the adopted half velocity width, $W$/2, which in some cases arises from an extrapolation to the rotation curve or from a 21 cm width (see Table 2); the vertical line is at $R_{\rm opt}$, the radius containing 83\% of the $I$ band flux.  A fit to the rotation curve is indicated by a solid line.  Note that the rotation curves are {\it not} deprojected to an edge--on orientation. \label{rcs}}
\end{figure}

In Figure \ref{rcs}, we plot the rotation curves folded about a kinematic center.  Only a portion of the full sample of rotation curves are presented here; the remainder of the rotation curves can be obtained by contacting the first author.  The horizontal dashed line in each panel indicates the adopted (and uncorrected for inclination) half velocity width, $W$/2, for each galaxy and the vertical dashed line is drawn at \Ropt.  We have overlayed fits to the rotation curves to facilitate, if necessary, shape corrections to the velocity widths.  Details on the fitting procedure can be found in G99.  We do not present a sampling of surface brightness profiles here; examples of surface brightness profiles can be found in Papers I and II, and any of the profiles can be obtained by contacting the first author.

\begin{figure}[!ht]
\centerline{\psfig{figure=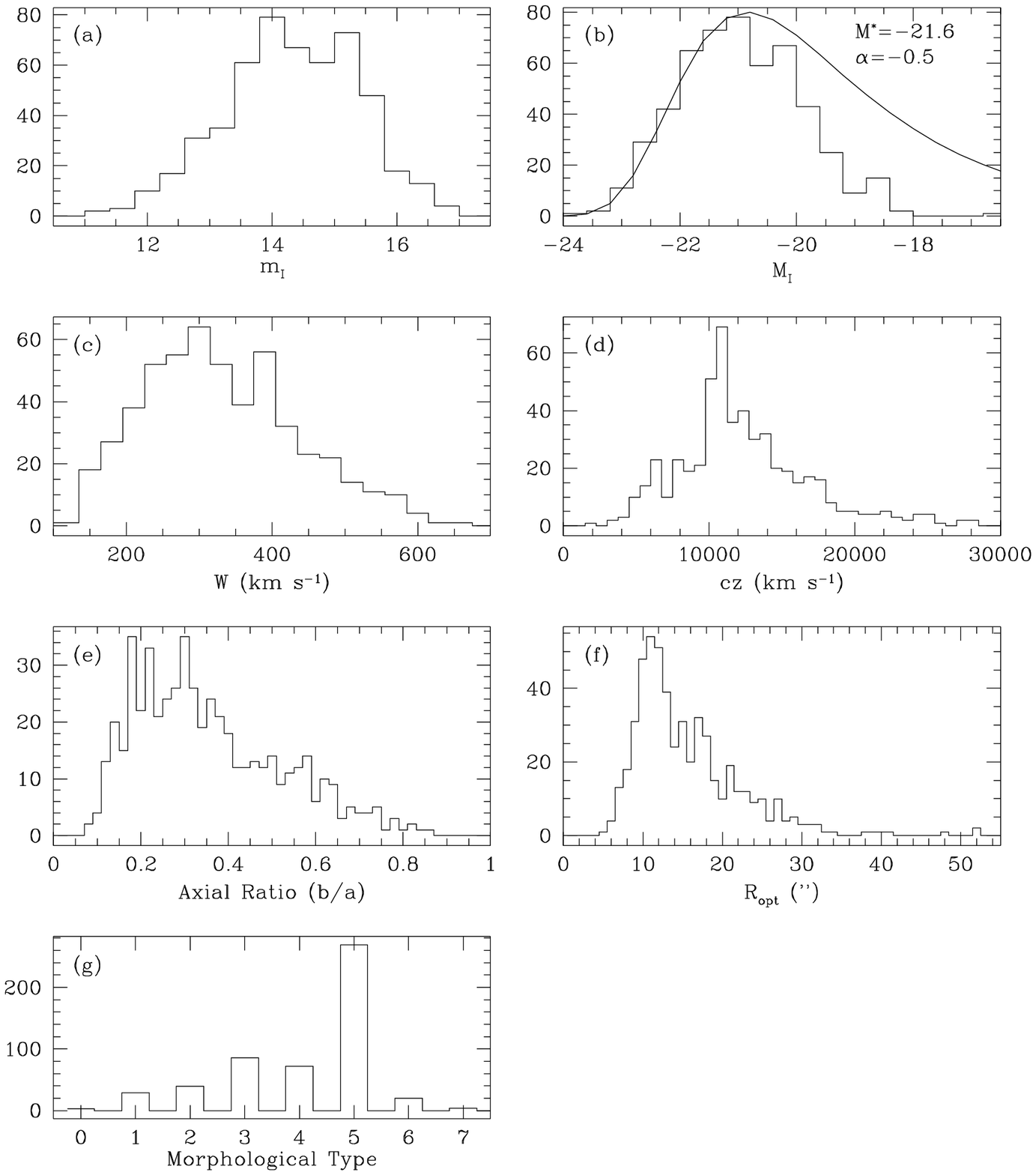,width=4in,bbllx=59pt,bblly=140pt,bburx=564pt,bbury=709pt}}
\caption[Sample Characteristics]
{\ Global distributions for the galaxy data from the entire 52 cluster sample are displayed.  From left to right, top to bottom: $I$ band apparent magnitude, $I$ band absolute magnitude (with Schechter function superimposed), rotational velocity width, recessional velocity in CMB frame, ratio of semi-minor axis to semi-major axis, optical radius, and morphological type in the RC3 scheme (de Vaucouleurs \etal\ 1991).}
\label{fig:hists}
\end{figure}

We present some general characteristics of the data for all 52 clusters in Figure \ref{fig:hists}.  Panels (a) and (b) are the apparent and absolute magnitude distributions in the $I$ band, with a Schechter LF ($M^*=-21.6, \alpha=-0.5$) overlayed in panel (b) for reference.  The full rotational velocity widths are given in panel (c), while panel (d) plots the redshift distribution of the entire data sample, foreground and background objects included.  We use the axial ratios in panel (e) to infer disk inclinations, and the \Ropt\ values from panel (f) are used for estimating velocity widths.  The last plot, panel (g), shows the morphological distribution of the galaxy sample.  The average CMB redshift of all the SCII galaxies ($N$=522) is 12,420 \kms, whereas the mean value excluding cluster foreground and background objects ($N$=441) is 11,890 \kms.

\begin{figure}[!ht]
\centerline{\psfig{figure=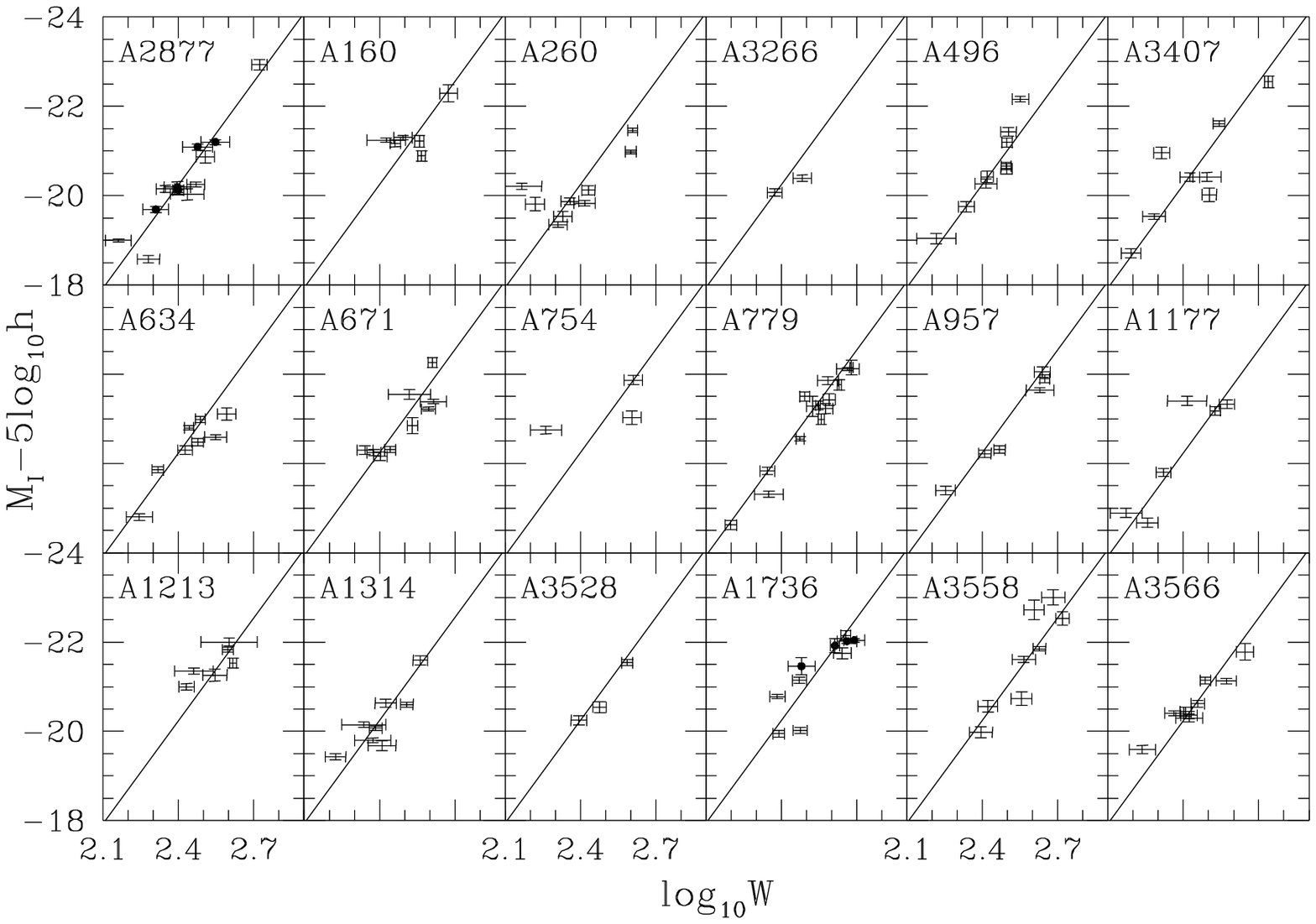,width=5.5in,bbllx=52pt,bblly=332pt,bburx=570pt,bbury=700pt}}
\caption[Cluster Tully-Fisher Diagrams]
{\ ``Raw'' TF plots for the 35 clusters are given.  The data have {\it not} been corrected for incompleteness bias.  In the A2877 and A1736 panels, the error bars containing filled circles represent members of ``A2877b'' and ``A1736b,'' clusters at higher redshifts than their conventional counterparts.  The dashed line is the template relation valid for low $z$ clusters, Equation \ref{eq:G97}. \label{tf}}
\label{fig:TFraw}
\centerline{Figure \ref{fig:TFraw} (Continued)}
\centerline{\psfig{figure=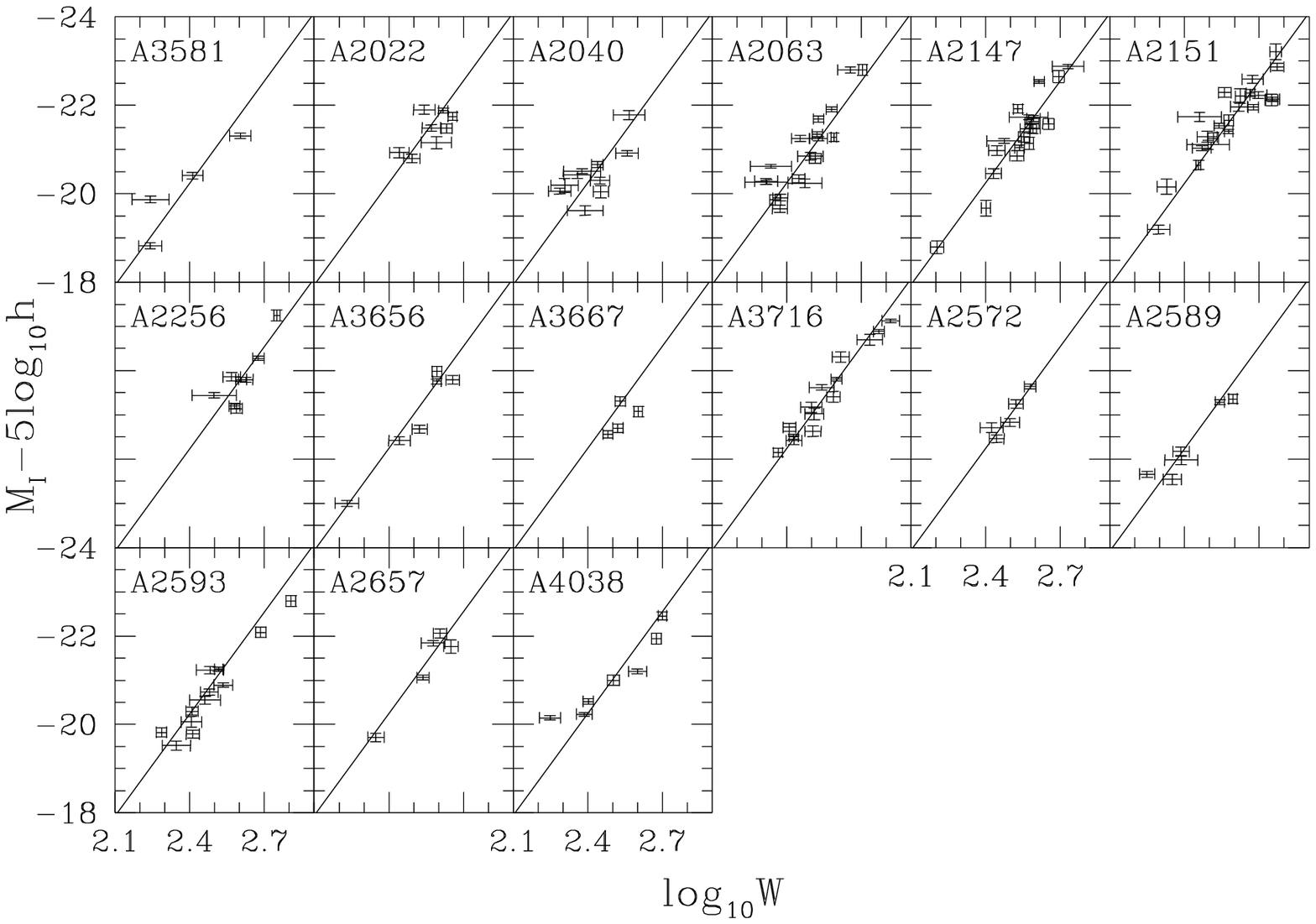,width=5.5in,bbllx=52pt,bblly=332pt,bburx=570pt,bbury=700pt}}
\end{figure}

Figure \ref{tf} gives the ``raw'' TF plots of each cluster uncorrected for any cluster incompleteness bias or peculiar motion.  A computation of such bias is presented in Paper V.  In the A2877 and A1736 panels, the error bars containing filled circles represent members of ``A2877b'' and ``A1736b,'' respectively.  Included in the TF plots, for reference, is the template relation obtained from the nearby sample of G97b:
\be
y = -7.68x - 21.01
\label{eq:G97}
\ee
where $y$ is $M_{\rm cor}$ -- 5log$h$ and $x$ is log$W_{\rm cor}$ -- 2.5.

\acknowledgements
We thank Katie Jore for the use of her rotation curve fitting programs.  The results presented here are based on observations carried out at the Palomar Observatory (PO), at the Kitt Peak National Observatory (KPNO), at the Cerro Tololo Inter--American Observatory (CTIO), and the Arecibo Observatory, which is part of the National Astronomy and Ionosphere Center (NAIC).  KPNO and CTIO are operated by Associated Universities for Research in Astronomy and NAIC is operated by Cornell University, all under cooperative agreements with the National Science Foundation.  The Hale telescope at the PO is operated by the California Institute of Technology under a cooperative agreement with Cornell University and the Jet Propulsion Laboratory.  The data reduction utilized IRAF procedures and a suite of programs developed at Cornell University.  IRAF is distributed by the National Optical Astronomy Observatories.  This research was supported by NSF grants AST94-20505 and AST96--17069 to RG and AST95-28960 to MH.  LEC was 
partially supported by FONDECYT grant \#1970735.

\newpage

\begin{thebibliography}{}

\bibitem[Abe 1989]{A89}
Abell, G., Corwin, H.G. \& Olowin, R.P. 1989, \apjs, 70, 1

\bibitem[Bra 1996]{B96}
Branchini, E., Plionis, M. \& Sciama, D.W. 1996, \apjl, 461, L17
 
\bibitem[]{}
Chiba, M. \& Yoshii, Y. 1995, \apj, 442, 82 

\bibitem[Cou 1993]{C93}
Courteau, S., Faber, S., Dressler, A. \& Willick, J. 1993, \apj, 412, L51


\bibitem[]{}
Courteau, S. \& Rix, H.-W. 1999, \apj, 513, 561

\bibitem[Dal 1997]{D97}
Dale, D., Giovanelli, R., Haynes, M., Scodeggio, M., Hardy, E. \& Campusano, L.
1997, \aj, 114, 455 [Paper I]

\bibitem[Dal 1998]{D98}
Dale, D., Giovanelli, R., Haynes, M., Scodeggio, M., Hardy, E. \& Campusano, L.
1998, \aj, 115, 418 [Paper II]

\bibitem[]{}
Dale, D., Giovanelli, R., Haynes, M., Campusano, L., Hardy, E. \& Borgani, S. 1999, \apjl, 510, L11 [Paper III]

\bibitem[]{}
Dale, D., Giovanelli, R., Haynes, M., Campusano, L. \& Hardy, E. 1999, \aj\  ({\it this edition}) [Paper V]

\bibitem[deV 1991]{d91}
de Vaucouleurs, G., de Vaucouleurs, A., Corwin, H. G., Buta, R. J., Paturel, G.
\& Fouque, P. 1991, {\it Third Reference Catalogue of Bright Galaxies}, (New York:Springer)

\bibitem[]{}
Feldman, H.A. \& Watkins, R. 1994, \apjl, 430, L17

\bibitem[Gio 1997a]{G97a}
Giovanelli, R., Haynes, M.P., Herter, T., Vogt, N.P., Wegner, G., Salzer, J.J., da Costa, L.N. \& Freudling, W. 1997a, \aj, 113, 22 [G97a]

\bibitem[Gio 1997b]{G97b}
Giovanelli, R., Haynes, M.P., Herter, T., Vogt, N.P., da Costa, L.N., Freudling, W., Salzer, J.J. \& Wegner, G. 1997b, \aj, 113, 53 [G97b]

\bibitem[]{}
Giovanelli, R., Haynes, M.P., Salzer, J J., Wegner, G., Da Costa, L.N. \& 
Freudling, W. 1998a, \aj, 116, 2632
 
\bibitem[]{}
Giovanelli, R., Haynes, M.P., Freudling, W., Da Costa, L.N., Salzer, J J. \& Wegner, G. 1998b, \apjl, 505, L91

\bibitem[]{}
Giovanelli, R. 1998c, in {\it Wide Field Survys in Cosmology}, ed. S. Colombi, Y. Mellier \& B. Raban (Paris: Editions Frontieres)

\bibitem[]{}
Giovanelli, R., Dale, D.A., Haynes, M.P. \& Hardy, E. 1999, submitted to \aj\ [G99]

\bibitem[]{}
Gramann, M., Bahcall, N.A., Cen, R. \& Gott, J.R. 1995, \apj, 441, 449

\bibitem[]{}
Haynes, M.P., Giovanelli, R., Herter, T., Vogt, N.P., Freudling, W.F., Maia, M.A., Salzer, J.J., Wegner, G. 1997, \aj, 113, 1197

\bibitem[Hud 1999]{H99}
Hudson, M.J., Smith, R.J., Lucey, J.R., Schlegel, D.J. \& Davies, R.L. 1999, \apjl, 512, L79

\bibitem[]{}
Karachentsev, I. 1989, \aj, 97, 1566

\bibitem[Kog 1993]{K93}
Kogut, A. \etal\ 1993, \apj, 419, 1

\bibitem[Lau 1994]{L94}
Lauer, T. \& Postman, M. 1994, \apj, 425, 418
 
\bibitem[]{}
Mathewson, D.S., Ford, V.L. \& Buckhorn, M. 1992, \apjs, 81, 413

\bibitem[]{}
Moscardini, L., Branchini, E., Tini--Brunozzi, P., Borgani, Plionis, M. \& Coles, P. 1996, \mnras, 282, 384

\bibitem[]{}
M\"{u}ller, K.R., Freudling, W., Watkins, R. \& Wegner, G. 1998, \apjl, 507, L105

\bibitem[]{}
Persic, M., Salucci, P. \& Stel, F. 1996, \mnras, 283, 1102

\bibitem[Rie 1995]{R95}
Riess, A., Press, W. \& Kirshner, R. 1995, \apjl, 445, L91

\bibitem[Sca 1992]{S92}
Scaramella, R., Vettolani, G. \& Zamorani, G. 1994, \apj, 422, 1

\bibitem[]{}
Strauss, M.A., Cen, R., Ostriker, J.P., Lauer, T.R. \& Postman, M. 1995, \apj, 444, 507

\bibitem[Tin 1995]{T95}
Tini-Brunozzi, P., Borgani, S., Plionis, M., Moscardini, L. \& Coles, P. 1995,
\mnras, 277, 1210

\bibitem[Wat 1996]{W96}
Watkins, R. \& Feldman, H. 1995, \apjl, 453, L73

\bibitem[Wil 1999]{W99}
Willick, J.A. 1999, \apj, 516, 47

\end{thebibliography}
\end{document}